# Succinct title: A PMT-like high gain avalanche photodiode based on GaN/AlN periodical stacked structure

# Running title: A PMT-like avalanche photodiode

Ji-yuan Zheng[1], Lai Wang[1], Di Yang[1], Jia-dong Yu[1], Xiao Meng[1], Yan-xiong E[1], Chao Wu[1], Zhi-biao Hao[1], Chang-zheng Sun[1], Bing Xiong[1], Yi Luo[1], Yan-jian Han[1], Jian Wang[1], Hong-tao Li[1], Julien Brault[2], Samuel Matta[2], Mohamed Al Khalfioui[2], Jian-chang Yan[3], Tong-bo Wei[3], Yun Zhang[3], Jun-xi Wang[3]

[1]Tsinghua National Laboratory for Information Science and Technology, Department of Electronic Engineering, Tsinghua University, Beijing, China.

[2]CNRS-CRHEA, Centre de Recherche sur L'Hétéro-Epitaxie et ses Applications- Centre National de la Recherche Scientifique, Valbonne, France.

[3]Institute of Semiconductors, Chinese Academy of Sciences, Beijing, China.

The email addresses of all authors:

| J. Y. zheng | zhengjiyuan11@mails.tsinghua.edu.cn | L. Wang | wanglai@mail.tsinghua.edu.cn |
|---|---|---|---|
| D. Yang | yang-d11@mails.tsinghua.edu.cn | J. D. Yu | yjd13@mails.tsinghua.edu.cn |
| X. Meng | mengx13@mails.tsinghua.edu.cn | Y. X. E | eyx11@mails.tsinghua.edu.cn |
| C. Wu | chao-wu12@mails.tsinghua.edu.cn | Z. B. Hao | zbhao@mail.tsinghua.edu.cn |
| C. Z. Sun | czsun@mail.tsinghua.edu.cn | B. Xiong | bxiong@mail.tsinghua.edu.cn |
| Y. Luo | luoy@mail.tsinghua.edu.cn | Y. J. Han | yjhan@mail.tsinghua.edu.cn |
| J. Wang | wangjian@mail.tsinghua.edu.cn | H. T. Li | lihongtao@mail.tsinghua.edu.cn |
| J. Brault | Julien.Brault@crhea.cnrs.fr | S. Matta | sm@crhea.cnrs.fr |
| M. Al Khalfioui | mak@crhea.cnrs.fr | J. C. Yan | yanjc@semi.ac.cn |
| T. B. Wei | tbwei@semi.ac.cn | Y. Zhang | yzhang34@semi.ac.cn |
| J. X. Wang | jxwang@semi.ac.cn | | |

Full contact details of the corresponding author:

L. Wang, Telephone: +86 18910557287; Fax number: 62784900; Email: wanglai@mail.tsinghua.edu.cn;

Y. Luo, Telephone: +86 13701153619; Fax number: 62784900; Email: luoy@mail.tsinghua.edu.cn

**Abstract**:

Avalanche photodiode (APD) has been intensively investigated as a promising candidate to replace the bulky and fragile photomultiplier tubes (PMT) for weak light detection. However, intrinsic limits in semiconductors make the former still inferior to the latter on device performance up to now. In conventional APDs, a large portion of carrier energy drawn from the electric field is thermalized, and the multiplication efficiencies of electron and hole are low and close. In order to achieve high gain, the device has to work under breakdown bias, wherein carrier multiplication proceeds bi-directionally to form a positive feedback multiplication circle. In this case, APDs should work under Geiger mode as a compromise between sustainable detection and high gain. On the other hand, PMT can achieve stable high gain under constant bias (linear mode). Here, we demonstrate an APD works like a PMT, which means it can work under constant bias and holds high gain without breakdown simultaneously. The device is based on a GaN/AlN periodically-stacked-structure (PSS). For the PSS holds the intrinsic features that there are deep Γ valleys and larger band offset in conduction band, electron encountered much less scatterings during transport in PSS APD. Electron holds much higher efficiency than hole to draw energy from the electric field, and avalanche happens uni-directionally with high efficiency. Extremely high ionization coefficient ($3.96 \times 10^5$ /cm) of electron and large ionization coefficient ratio (over 100) between electron and hole is calculated in the PSS APD by Monte-Carlo simulations and a recorded high gain ($10^4$) tested under constant bias without breakdown is obtained in a prototype device, wherein the stable gain can be determined by the periodicity of the GaN/AlN PSS and no quenching circuits are needed for sustainable detection. This work not only brings a new light into avalanche multiplication mechanism, but also paves a technological path to realize highly sensitive APD working under constant bias like PMT.

**Introduction:**

The discovery of photoelectric effect by Hertz in the late 19th century has played an important role in establishing the quantum theory. Einstein's introduction of photon theory successfully explained the photoelectric effect. Since then, the optoelectronic technology has met with its dramatic development[1]. In 1890s, Elster and Geitel invented the first phototube, in which the current response to illumination is detected. The phototube is considered as one of the earliest electronic tubes and has played an important role in scientific research and industry[1, 2]. Many improvements have been carried out on phototube, and the most important improvement is the introduction of multiplication ability. In 1930s, photomultiplier (PMT), a kind of unipolar multiplication phototube, was developed[3, 4]. In PMT, photons are absorbed by photocathode and then excite the electrons to vacuum levels. The electrons then fly to the accelerating region surrounded by dynode arrays and gather kinetic energy high enough to trigger ionizations in dynodes. The secondary electrons again accelerate and trigger new round of ionization in the next dynode. This progress successively happens and a greatly multiplied number of electrons are collected by anode afterward[1]. Generally, a high electric field and accelerating distance product is needed to enhance the multiplication effects. However, positive ions could also be generated from residual gas present in the tube under high electric field. The positive ions and electrons drift toward two opposite directions, leading to uncontrolled regeneration[5]. Thus, to make sure only electrons fly between dynodes and trigger multiplication, high vacuum and relatively long accelerating distance are needed in fabricating the device. Up to now, PMT is the most successful weak signal detectors for its high gain (greater than $10^6$) and constant bias (linear-mode) operation.

But the PMT should be made as large and fragile vacuum tube to achieve the extraordinary performances, which limits its applications in many frontier domains.

Replacing the vacuum tube device by solid state device is the megatrend of electronic engineering and great successes have been achieved on replacing vacuum tubes by semiconductor devices. For example, transistor and integrated circuit chip have replaced electronic valve and create current microelectronics industry; light emitting diode (LED) has also been developed as a replacement of traditional filament lamp or fluorescent lamp. However, as with the photo detector for weak optical signal, solid state device still cannot replace the PMT thoroughly.

Avalanche photodiode (APD) has been intensively investigated as a promising candidate to replace PMT for weak light detection[6-9]. However, as shown in Figure 1a, in an APD, there are two kinds of carriers, electrons and holes. Both of them can trigger the impact ionization to generate new electron-hole pairs. As large quantities of scatterings accompany with carriers' transport, most of the energy carriers drawing from electric field is thermalized. As a result, the ionization coefficients (reciprocal of the mean distance between two successive impact ionization events) for carriers are not high[10-15]. Under a low electric field, the difference between electron ionization coefficient and hole ionization coefficient is large, which can be regarded as that only one kind of carrier is permitted to trigger ionizations. In this case, even though the APD can work like a PMT, it is hard to achieve high multiplication gain due to the low ionization efficiency. For example, the electron's ionization coefficient is only about $2\times10^4$ /cm for silicon when the ionization coefficient ratio between electron and hole is about 10 (ref. 16, 17); the hole's ionization coefficient is only about $2\times10^3$ /cm for GaN when the ionization coefficient ratio between hole and electron is about 10 (ref.14). As a result, the highest gain for silicon APDs working under constant bias is only about 100 (ref. 18) and that for GaN APDs is only about 10 (ref. 19). Therefore, as a compromise way to achieve high multiplication gain, conventional HJ APD should work under extremely high bias, wherein ionization coefficients of electron and hole become close. In other words, once there generates new electron-hole pair, both electron and hole will trigger ionizations. As a result, as shown in Figure 1b, the multiplication proceeds along two opposite directions and form positive feedback chains, then the response current drastically rises, which is often called "breakdown". Since the bipolar carrier ionization would not stop until the bias voltage is reduced below its breakdown value. In practical applications, APDs need to be periodically quenched below breakdown bias to avoid permanent damage, and photons can be detected only when APDs are under breakdown. This compromise between high gain and sustainable detection, commonly known as Geiger-mode[20, 21], restricts the applications of current APDs for the complexity of system. Since the reasons mentioned above are intrinsic, these crucial bottlenecks have perplexed the researcher for many years. The most ideal APD should work like a PMT, wherein only one kind of carriers can trigger the impact ionization, the avalanche happens uni-directionally without breakdown as shown in Figure 1c, and meanwhile the ionization coefficient must be ultrahigh to achieve a high gain.

Many attempts were made to improve the efficiency and controllability of the multiplication process in APDs. For example, F. Capasso *et al*. proposed in 1982 (ref. 22) an APD based on AlGaAs/GaAs superlattice structure, wherein the large band offset ratio of conduction band (CB) to valance band (VB) could help to increase the ratio of ionization efficiencies between electrons and holes. However, it was shown by experiments and Monte Carlo simulations that this increase is quite limited because the energy drawing process of electrons is greatly interfered by thermalization[23]. Consequently, high electric field is still necessary for electrons to overcome thermalization and trigger ionization. Unfortunately, the holes also have opportunity to trigger ionization under such strong electric field. Thus, the multiplication process in the AlGaAs/GaAs superlattice APD is still inefficient. Thereafter, many improvements are brought out based on impact ionization engineering ($I^2E$) method[24-28]. By inserting doping layers or heterojunction into each cycle of the superlattice, the ionization coefficient ratio can be enhanced at some degree. However, since scattering problems are unsolved yet, the ionization coefficient is still not high. For example, the electron's ionization coefficient is about $5 \times 10^4$ /cm for $I^2E$ designed APD based on InGaAs and InAlAs when the ionization coefficient ratio between electron and hole is about 25 (ref. 24).

Essentially, the major challenge in achieving a PMT like APD is to keep one kind of carrier drawing energy from the electric field more efficiently than the other with much less thermalization rate. III-nitride semiconductors, specifically GaN and AlN widely used due to their excellent physical and chemical properties[29, 30], are promising media to achieve this goal.

In conventional HJ and $I^2E$ designed APD based on most commonly used materials, the deepness of independent valleys in conduction band or valence band are not higher than the bandgap, so intervalley scatterings will frequently impact the carrier transport before it achieves the ionization threshold energy[14], resulting in a low ionization coefficient. On the other hand, there is a deep Γ valley (about 2 eV) in the first CB of both GaN and AlN with extremely low density of states (DOS), while no such valleys exist in the VB[31, 32]. Electrons transporting within the Γ valley of either GaN or AlN would experience much weaker thermalization. Although the depth of Γ valley is not larger than the ionization threshold energy for electrons (5.3 eV) in GaN[33], there is a 2-eV CB offset between GaN and AlN[34]. Electrons transporting from GaN to AlN along (0001) direction will return back to the Γ valley of AlN and continue their transport with weak thermalization until their energy rises to higher than 4.0 eV, which is much closer to the ionization threshold energy (5.3 eV). This means that the electron energy could be raised to a value higher than the ionization threshold with only a moderate electric field. Accordingly, we propose an APD with GaN/AlN periodically stacked structure (PSS) as shown in Figure 1d to realize controllable electron multiplication.

The working principle of the GaN/AlN PSS APD can be described as follows: Electron generated by UV light drifts into multiplication region. First, it draws energy in the Γ valley of GaN with ignorable thermalization during acceleration until its energy reaches 2 eV. Then, it comes into the Γ valley of AlN and continues drawing energy from electric field. When it leaves AlN layer and enters into the next GaN layer, it has gathered enough energy to trigger ionization and generate new electron-hole pairs. The electrons can repeat the above process in subsequent periods of GaN/AlN. On the other hand, the ionization probability for hole is still greatly restricted, since there are no deep separated valleys in valence band and hole transport is intensively interfered by thermalization in either GaN or AlN. Hence, GaN/AlN PSS APD can realize large difference in ionization coefficients between electrons and holes, which means the avalanche could occurs uni-directionally and is also controllable.

## Materials and methods:

### Ensemble Monte Carlo simulations

Ensemble Monte Carlo simulations based on first-principle theory are carried out for conventional GaN-based HJ APD and our proposed GaN/AlN PSS APD. Firstly, the energy band structure and wave functions of GaN and AlN are calculated. It is noted that there are many different views on the detailed band structure of GaN[31, 32, 35-37]. Here, we take the commonly used empirical pseudopotential method to calculate the full band structure[31, 32]. The Schrödinger equation is solved to obtain the electron states[37]. Fifteen groups of pseudo-potential parameters in relative super symmetrical points are used to solve the equations[14]. To expand the periodic part of the Bloch wave-function, 183 basic sets of reciprocal lattice vectors are used. Then, scattering and ionization rates of GaN and AlN are calculated using Fermi Golden Rules. For simplification, only two dominant normal scattering mechanisms, i.e. deformation potential scattering and polar optical phonon scattering, are included. The model can be referred to ref. 14, 38, 39. Finally, Monte Carlo simulations are carried out. The trajectory of a single carrier is monitored as it travels under the influence of the electric field and random scattering. The statistical behavior of carrier transport can be described by the averaged transport behavior of this carrier over plenty of its trajectories. For GaN-based HJ APD, the simulation model is described in previous studies[14, 32, 38, 39]. For GaN/AlN PSS APD, we take the following assumptions: (1) for simplification, electric field is regarded identical in both AlN and GaN layers without the consideration of polarization field; (2) interface scattering is ignored; (3) energy conservation is fully ensured when carriers are getting through the GaN/AlN interfaces; (4) momentum conservation is ensured by selecting a final state with a momentum closest to that of the initial state and (5) if the energy of carrier flying from GaN to AlN is not higher than the band offset, an energy-state reset is put on the carrier and the spatial position is reset to the origin of the AlN layer.

The structure of GaN/AlN PSS is taken as endless cycles of periodic GaN (10 nm)/AlN (10 nm). In the simulation, a single electron is driven by an electric field of 3.2 MV/cm along (0001) direction. The trace of the electron is sampled at a time interval of 0.5 fs, and both its energy state and position are recorded.

**Growth and fabrication of GaN/AlN PSS APD**

For further demonstration, a prototype of GaN/AlN PSS APD is fabricated. The epitaxial structure of this device is plotted in Figure 2a. The framework of epi-layer is GaN-based p-i-p-i-n separated absorption and multiplication structure[40] and the multiplication layer is replaced with a 20-period GaN (10 nm)/AlN (10 nm) PSS. The samples were grown on 2-inch (0001) c-plane sapphire substrates. The heterostructure initially consists of a 4-μm thick AlN template from Suzhou Nanowin Science and Technology Co., Ltd. Then, the sample was introduced in a RIBER 32 P reactor for growth by molecular beam epitaxy. The fabrication process included the growth of a thin (~ 50 nm) AlN layer on top of the template. Next, a Si-doped 500 nm-thick GaN layer was grown, corresponding to an n-type doped layer with a carrier concentration ~ $4 \times 10^{18}$ cm$^{-3}$. The periodically stacked structure (PSS) was then fabricated with an identical GaN and AlN layer thickness of 10 nm and a stack of 20 layers. After the PSS growth, a Mg-doped 10 nm-thick GaN layer was grown, followed by the growth of a 300 nm-thick GaN non-intensively-doped layer. Finally, a 100 nm-thick Mg-doped GaN was deposited. The Mg doping growth conditions correspond to a p-type doped layer with a carrier concentration ~ $1 \times 10^{18}$ cm$^{-3}$. The cross sectional transmission electron microscope (TEM) photo of the epitaxial layers is shown in Figure 2b. The thickness of each layer is uniform and coincides with the designed value.

A double-mesa structure (inner diameter 25 μm, outer diameter 35 μm) as shown in dashed circle region in Figure 2c is used in fabricating the device. The mechanism of double mesa can be referred in Ref. 41. The double-mesa is formed by inductively coupled plasma dry etching technique. The light absorption surface is in the 25-μm inner mesa. Plasma enhanced chemical vapor deposition is adopted to deposit SiO$_2$ passivation layers. Ni (2.5 nm)/Au (8 nm) and Cr (25 nm)/Au (200 nm) are used as the p-type transparent ohmic contact and n-type ohmic contact, respectively. Au (100 nm) is used as electrode pad. The top view micrograph of the fabricated device is shown in Figure 2d. The device fabrication is in high quality that the size and position of each component of the device are in full compliance with the design.

**Device measurement**

*I-V* curves are measured by an Agilent Type 4155C semiconductor analysis meter. First, reverse biased dark currents are measured with no light illuminating the chip. Then, the chip is front-illuminated by a xenon lamp with a monochromator to obtain a 350-nm wavelength signal. Input light power density is measured by standard power meter, of which the photosurface is round with a diameter of 1 cm (much less than the output light spot of the monochromator). The tested light power density is 51 µW/cm$^2$. Thus, the light power coupled into the chip is 250 pW calculated by geometrical relationship, wherein the absorption area is taken as the 25 µm inner mesa.

In traditional APDs, ratio of the photocurrent $I_r$ (difference between light current $I_l$ and dark current $I_d$) under high and unit-gain bias is used to characterize the multiplication ability of devices[8]. However, in PSS structure, there is no unit-gain region under low-bias in the photocurrent-voltage curve. We believe this is because the internal quantum efficiency of APD increases with bias and meanwhile the multiplication happens at very low bias. Therefore, external quantum efficiency (*EQE*), calculated as,

$$EQE = (I_r \times E_p) / (e \times P) \qquad (1)$$

is taken as a criterion to reflect the multiplication performance, where *e*, *P* and $E_P$ denote the elementary charge, the power of incident light and the photon energy, respectively. Hence $I_r$ is only 70.6 pA when *EQE* is 100%.

A temperature dependent test is carried out in the LakeShore Type CRX-4K cryogenic probe station. *I-V* curves and related *EQE-V* curves are tested under temperature of 270, 300 and 330 K.

**Results and discussion**

**Monte Carlo simulation results**

The calculated band structures of AlN and GaN within the first Brillouin zone are shown in Figures 3a and 3b, respectively, where the *E-k* relationships along supersymmetric directions within the entire Brillouin zone are depicted. A deep Γ valley (~2.0 eV) is evident in the first CB of both GaN and AlN. The corresponding DOS as a function of energy in the first Brillouin zone are plotted in Figures 3c and 3d, showing that the energy at the bottom of CB is 3.4 eV for GaN and 6.2 eV for AlN, respectively (as a reference, energy at the top of VB is defined as 0 eV). Hence, the DOS is extremely low only within Γ valleys. As the phonon scattering rate is positively correlated with DOS[14], electron transporting in Γ valley of first conduction band meet much less thermalization than transporting outside the valley. Moreover, as the DOS for valence band increases fast in low energy region, the thermalization for hole transport is much more serious. It can be predicted that with the aid of 2-eV conduction band-offset, electrons could trigger ionization much easier than holes.

As a vivid presentation of the transport process for electrons, the energy distribution as a function of position along (0001) direction in both structures calculated by Monte-Carlo simulations are plotted in Figures 4a and 4b, respectively. In GaN HJ, the electron energy increases relatively fast when it is lower than 2.0 eV, then becomes greatly constrained by scattering while most of the kinetic energy drawn from the electric field is thermalized. The energy vibrates drastically between 2 eV and 4 eV, which is much lower than the ionization threshold energy (5.3 eV). Although the energy could overshoot and reach 5.3 eV, the ionization is still a low probability event. This picture is similar to holes (not shown). As a result, the multiplication region should be thick enough or higher electric field is required to trigger bipolar ionization of electrons and holes[42]. In contrast, in GaN/AlN PSS, electrons draw energy quite drastically until 4.0 eV. In addition, electrons can touch the highest energy level (around 6.0 eV) after it returns back to GaN from AlN along (0001) direction, which is sufficiently higher than the ionization threshold energy (5.3 eV) in GaN. The possibility for electrons not triggering ionization and being fast thermalized (as shown in Figure 4b) is quite low. In other words, the ionization process is highly efficient and the total number of ionization events can be controlled by the period number of PSS. Furthermore, holes in the PSS APD still experience intensive scattering even though there is a 0.7 eV VB offset between AlN and GaN. Specifically, ionization coefficients in GaN/AlN PSS APDs and conventional GaN HJ APDs as a function of electric field are calculated and shown in Figure 4c. It is clear that the ionization coefficient for electrons is drastically improved in the former, while the improvement in ionization coefficient for the holes is negligible. For electric field stronger than 2.8 MV/cm, electron ionization reaches a saturated value around $3.96 \times 10^5$ /cm. To our best knowledge, such high ionization coefficient has never been realized in a HJ APD or I$^2$E designed APD before[16, 17, 26]. The miraculous improvement in ionization coefficient is due to the significant decrease of scatterings. The corresponding average ionization length of electrons approximately equals to 20-nm, or the period of GaN/AlN PSS. In other words, the ionization coefficient saturates when the electron-triggered ionization happens nearly once in a period with high efficiency.

**Device performance**

The *I-V* and *EQE-V* curves tested under room temperature are shown in Figure 5a. Unlike conventional APDs, the *EQE* of the PSS APD increases gradually with bias voltage. Moreover, the *EQE* saturates at a high level of $1 \times 10^4$ (~ $2^{14}$), i. e. the photon-generated electrons are repeatedly multiplied by at least 14 times (the average number of ionization events during a single-pass transport in PSS). Equivalently, electrons trigger ionization nearly once within a period for all the 20 periods.

It is noticed that the saturation phenomenon of *EQE* curve shown in Figure 5a has a similar trend with the electron ionization coefficient shown in Figure 4c, which rules out the possibility that this device is a kind of photoconductive detector[43]. Generally, multiplication phenomenon in a photovoltaic detector could be attributed to three mechanisms: thermal breakdown, Zener tunneling, and impact ionization [44]. All possibilities are taken into consideration and ruled out one by one until the essential mechanism is found out. Firstly, thermal breakdown is ruled out because the multiplication phenomenon is repeatable. In a thermal breakdown process, the material is destroyed and the breakdown can't be repeated[44]. Secondly, Zener tunneling is ruled out because the dark current increases with temperature as shown in Figure 5b, while Zener tunneling current decreases with increasing temperature[44]. Finally, the impact ionization is attributed to the multiplication while no microplasma avalanche or premature avalanche are observed as the *EQE* decreases with temperature, especially under high bias as shown in Figure 5c[44, 45]. In PSS APDs, although phonon scattering has been greatly reduced compared with conventional HJ APD, the rudimental scattering still holds impacts on the multiplication performance. For the scattering degree has a positive relationship with temperature, the ionization efficiency of electron and *EQE* decrease with the increase of temperature.

As a result of the principle mentioned above, some interesting characteristics of this device are found. Unlike conventional HJ APDs, the PSS APD could work under constant bias with high gain. As shown in Figure 6, current measured under a constant bias of 65 V and room temperature responses well with on/off switching of incident UV light, where the EQE reaches about $10^4$. Conventional APDs working under constant bias simply cannot reach such high gain[22,

[24, 25, 46, 47]. All these phenomena validate that the electron-triggered ionization in GaN/AlN PSS APD is highly-efficient and controllable. Furthermore, as the electron's ionization is extremely high and the multiplication performance is determined by periodicity, it can be expected that higher gain working under constant bias comparable with PMT can be achieved by simply increasing the cycles of PSS.

**Conclusions**

A GaN/AlN periodically-stacked-structure (PSS) is proposed and demonstrated to manipulate the carrier transport and realize a highly-efficient and controllable carrier multiplication process. An extremely high electron's ionization coefficient of $3.96 \times 10^5$ /cm and large ionization coefficient ratio (over 100) between electron and hole is calculated in the PSS APD by Monte-Carlo simulations. The avalanche multiplication process happens uni-directionally and results in a recorded high gain ($10^4$) under constant bias in a prototype device. It is the very first time that an APD could work under constant bias without breakdown and achieve a high gain comparable with Geiger Mode APD. The present work is just an example of UV APD. Devices with various working wavelengths ranging from infrared to ultraviolet can be realized utilizing the transition between bands or sub-bands by changing the absorption layer to other III-nitride semiconductors. Potential applications such as electronic amplifiers could also be developed by using this multiplication mechanism.

**Acknowledgment:** J.Y. Zheng, L. Wang, D. Yang, J. D. Yu, X. Meng, Y.X. E, C. Wu, Z.B. Hao, C.Z. Sun, B.Xiong, Y. Luo, Y.J. Han, J. Wang, H.T. Li would like to thank C.C. Fan for a detailed discussion of the manuscript. They are highly indebted to the National Basic Research Program of China (Grant Nos. 2012CB3155605, 2013CB632804, 2014CB340002 and 2015CB351900), the National Natural Science Foundation of China (Grant Nos. 61574082, 61210014, 61321004, 61307024, and 51561165012), the High Technology Research and Development Program of China (Grant No. 2015AA017101), the Independent Research Program of Tsinghua University (Grant No. 2013023Z09N), the Open Fund of the State Key Laboratory on Integrated Optoelectronics (Grant No. IOSKL2015KF10), and the CAEP Microsystem and THz Science and Technology Foundation (Grant No.CAEPMT201505). J. Brault, S. Matta and M. Al

Khalfioui acknowledge the support of this work by ANR Project <ANR-14-CE26-0025-01> NANOGANUV. They also would like to thank J. Y. Duboz for a critical reading of the manuscript.

Figure 1 The motivations of the new design: (a) Carrier transport in conventional HJ APDs under electric field, (b) Bipolar ionization proceeding along two opposite directions and (c) Unipolar ionization proceeding along a single direction. (d) Carrier transport in GaN/AlN PSS APDs.

Figure 2 Design and fabrication of the GaN/AlN PSS APD: (a) Epi-structures of the GaN/AlN PSS APD. (b) Cross sectional TEM image of the epitaxy layers. (c) Device processing design. (d) Top view micrograph of a practical device.

Figure 3 Material properties of GaN and AlN: (a) Energy band of GaN, (b) Energy band of AlN, (c) DOS in GaN and (d) DOS in AlN.

Figure 4 Transport properties for PSS APD and HJ APD. (a)&(b) Electron energy variation during its transport in GaN HJ APD (a) and GaN/AlN PSS APD (b) under an electric field of 3.2 MV/cm. (c) Ionization coefficients of electron and hole as a function of electric field in conventional GaN HJ APD and GaN/AlN PSS APD.

Figure 5 Measured results of GaN/AlN PSS APD. (a) *I-V* and *EQE-V* curves; (b) Reverse biased dark current curves tested under different temperatures. (c) Temperature dependence of *EQE-V* curves.

Figure 6 The device working under constant bias responses well with UV light switching on and off.

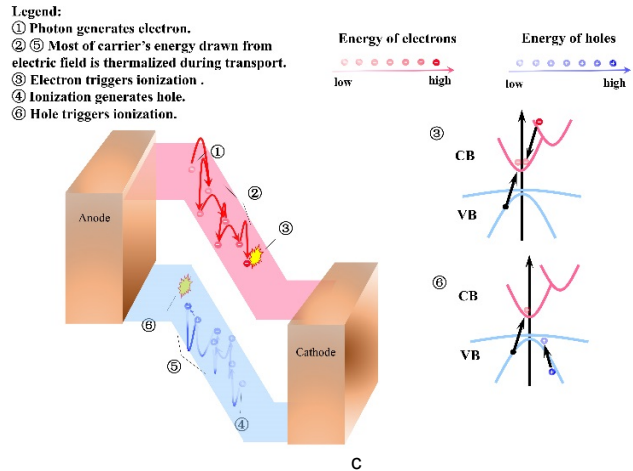
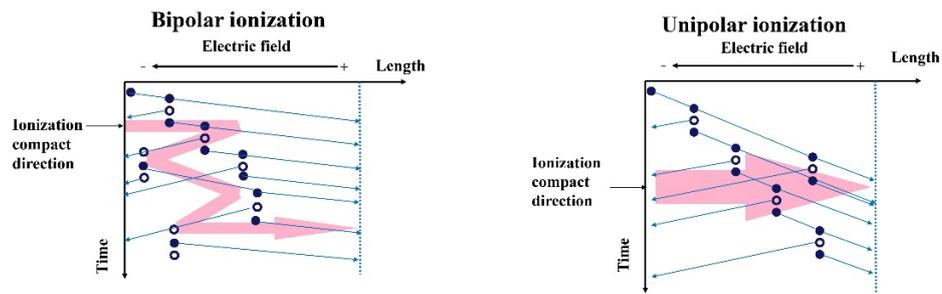
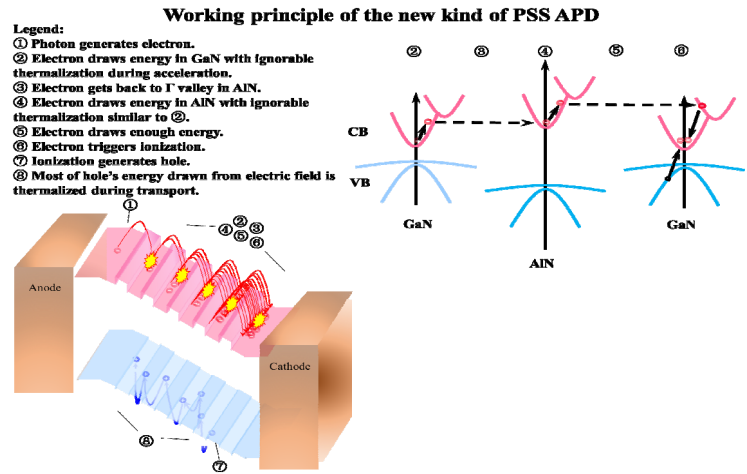

Figure 1 The motivations of the new design: (a) Carrier transport in conventional HJ APDs under electric field, (b) Bipolar ionization proceeding along two opposite directions and (c) Unipolar ionization proceeding along a single direction. (d) Carrier transport in GaN/AlN PSS APDs.

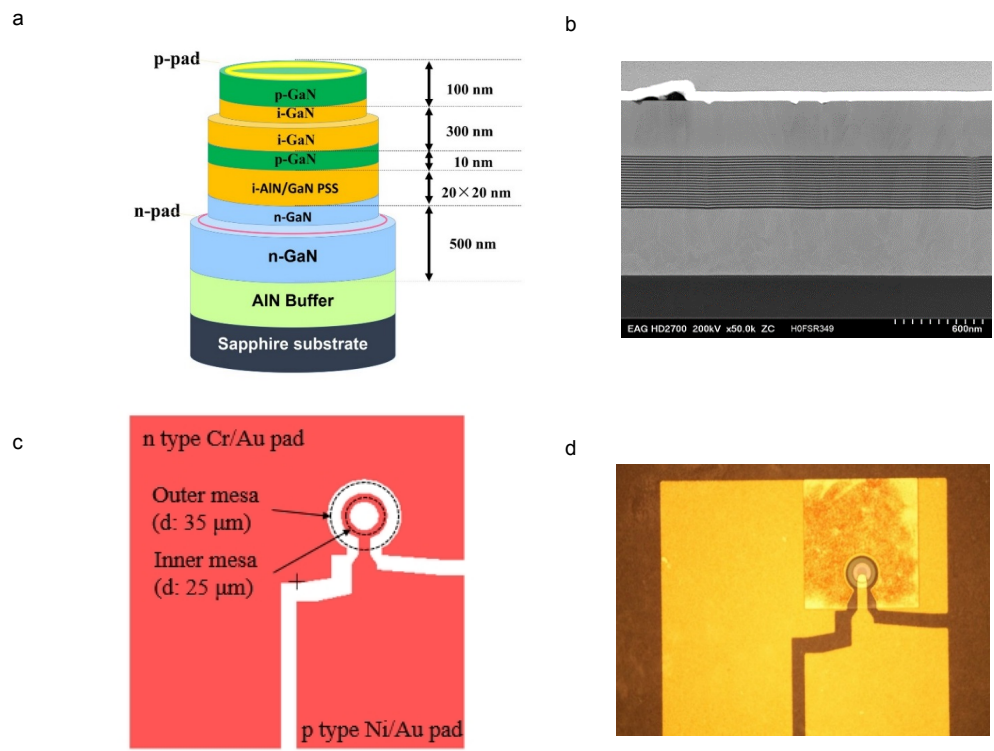

Figure 2 Design and fabrication of the GaN/AlN PSS APD: (a) Epi-structures of the GaN/AlN PSS APD. (b) Cross sectional TEM image of the epitaxy layers. (c) Device processing design. (d) Top view micrograph of a practical device.

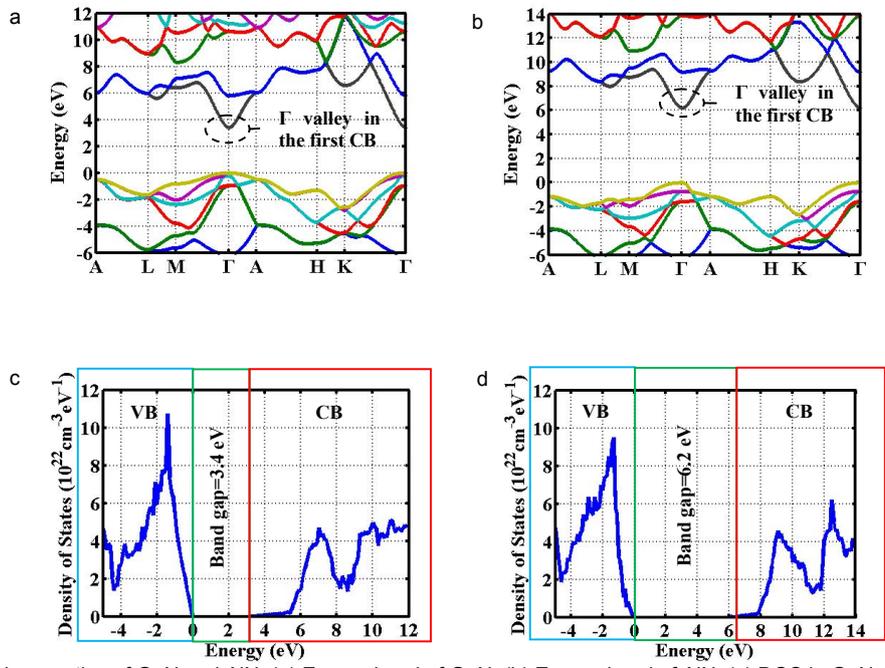
Figure 3 Material properties of GaN and AlN: (a) Energy band of GaN, (b) Energy band of AlN, (c) DOS in GaN and (d) DOS in AlN.

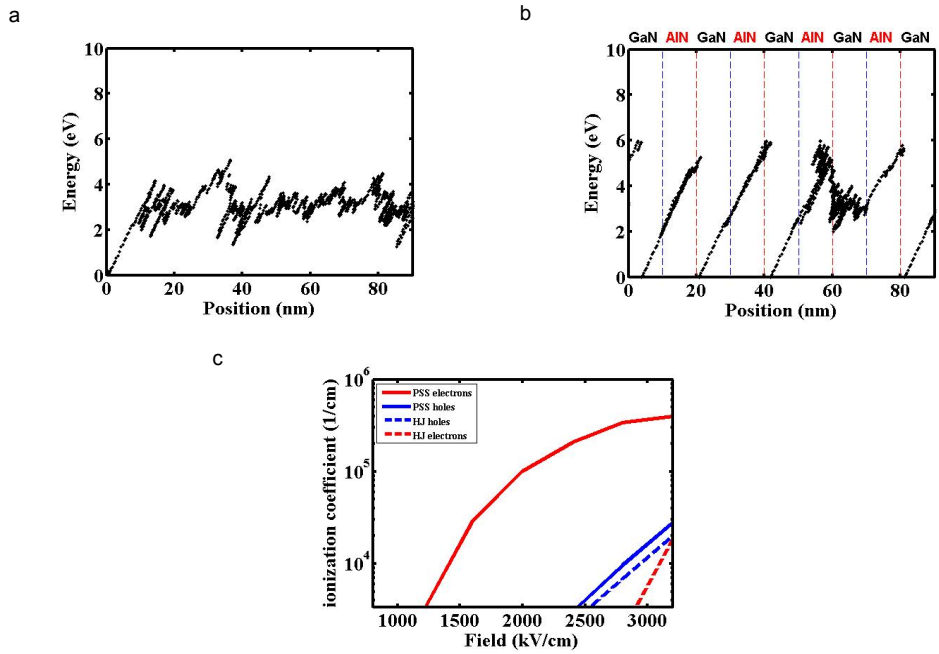

Figure 4 Transport properties for PSS APD and HJ APD. (a)&(b) Electron energy variation during its transport in GaN HJ APD (a) and GaN/AlN PSS APD (b) under an electric field of 3.2 MV/cm. (c) Ionization coefficients of electron and hole as a function of electric field in conventional GaN HJ APD and GaN/AlN PSS APD.

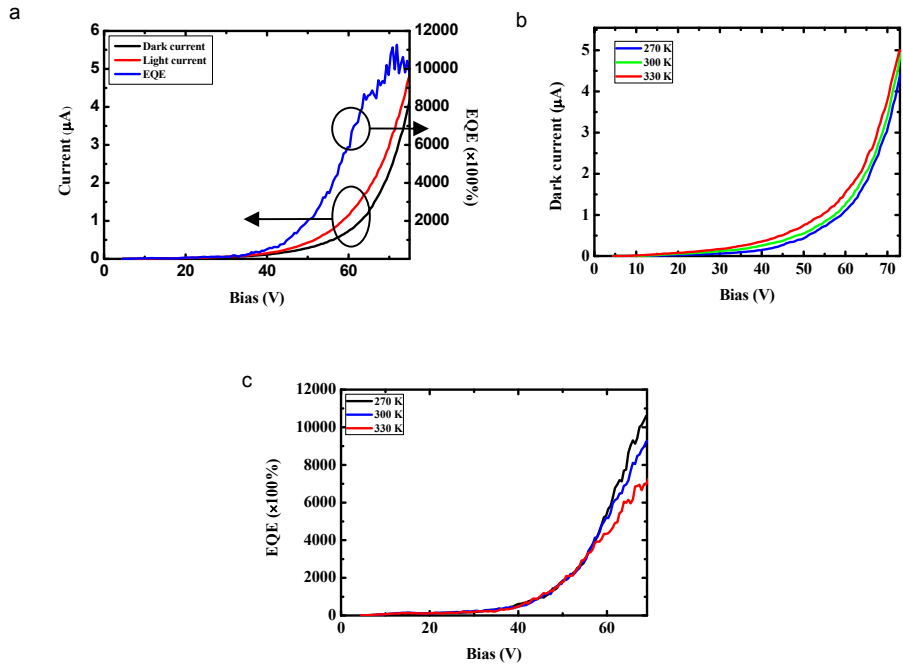

Figure 5 Measured results of GaN/AlN PSS APD. (a) *I-V* and *EQE-V* curves; (b) Reverse biased dark current curves tested under different temperatures. (c) Temperature dependence of EQE-V curves.

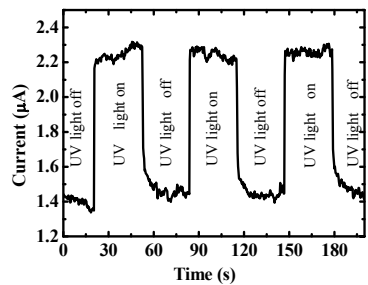

Figure 6 The device working under constant bias responses well with UV light switching on and off.